# Data Governance and Data Management in Operations and Supply Chain: A Literature Review

Xuejiao Li[1], Yang Cheng[1,*], Xiaoning Xia[1], Charles Møller[2]

1: Department of Materials and Production, Aalborg University, Denmark

2: Department of Mechanical and Production Engineering, Aarhus University, Denmark

* Corresponding author, cy@business.aau.dk, cy@mp.aau.dk

**Abstract:** In the dynamic landscape of contemporary business, the wave in data and technological advancements has directed companies toward embracing data-driven decision-making processes. Despite the vast potential that data holds for strategic insights and operational efficiencies, substantial challenges arise in the form of data issues. Recognizing these obstacles, the imperative for effective data governance (DG) becomes increasingly apparent. This research endeavors to bridge the gap in DG research within the Operations and Supply Chain Management (OSCM) domain through a comprehensive literature review. Initially, we redefine DG through a synthesis of existing definitions, complemented by insights gained from DG practices. Subsequently, we delineate the constituent elements of DG. Building upon this foundation, we develop an analytical framework to scrutinize the collected literature from the perspectives of both OSCM and DG. Beyond a retrospective analysis, this study provides insights for future research directions. Moreover, this study also makes a valuable contribution to the industry, as the insights gained from the literature are directly applicable to real-world scenarios.

# 1. Introduction

In the contemporary business landscape, the abundance of data and technologies has prompted companies to transition toward data-driven decision-making processes. The allure of leveraging data for strategic insights and operational efficiencies has driven organizations across industries to invest significantly in data collection and analytics. However, this journey towards data-driven decision-making is not without its challenges. While data holds immense potential, companies encounter formidable challenges in realizing this potential. Issues such as data quality, integrity, and security pose significant barriers to the seamless adoption of data-driven technologies and approaches. As companies grapple with these obstacles, the need for effective data governance (DG) becomes increasingly apparent. DG emerges as a strategic imperative, offering a structured framework to manage data-related challenges and pave the way for organizations to derive maximum value from their data assets.

DG stands out as a key framework to address the complexity of managing data within organizations. It encompasses policies, processes, and standards that ensure high data quality, facilitate proper data management, and unleash the potential value embedded in the vast sea of data (Power & Trope, 2006; Abraham et al., 2019; Jatin, 2023). While DG has been studied in the Information Systems (IS) domain, the significance of DG in the Operations and Supply Chain Management (OSCM) domain is largely unexplored in the current literature. The OSCM domain, characterized by its unique set of challenges and requirements, necessitates a dedicated exploration of DG tailored to its specific needs.

Building upon these motivations, this research endeavors to bridge the existing gaps by conducting a comprehensive literature review focused on DG within the OSCM domain. The goal is to synthesize and critically evaluate existing knowledge, examine the relevance and applicability of DG frameworks, methodologies, and best practices in the context of operations and supply chains (O&SC). Through this examination, the research aims to identify gaps, challenges, and opportunities specific to the OSCM domain related to DG, laying the groundwork for future studies, and developing practical guidelines for organizations in industry. By doing so, this study seeks to contribute to the development of a robust foundation for effective DG practices tailored to the unique challenges and dynamics in the O&SC context.

The structure of this study is as follows. Section 2 introduces the theoretical background, including the definition of DG and its analogous concept, data management (DM). Additionally, it presents an overview of the existing literature reviews on DG. Section 3 outlines the methodology, which includes the search strategy, statistical description of selected literature, and the framework for literature analysis. Section 4 presents the results of the literature analysis based on the framework proposed in Section 3. Section 5 highlights potential future research directions. Finally, Section 6 concludes the study and indicates the limitation.

## 2. Theoretical background

### 2.1 Definition of DG

Prior to undertaking the literature review, it is imperative to introduce the concept of DG. The discourse surrounding DG dated back to, at the very least, 2011. Table 1 presents five frequently referenced definitions of DG.

Table 1. Frequently referenced definitions of DG

| Researchers | Definition of data governance |
|---|---|
| **Otto (2011b)** | Data governance is a companywide framework for assigning decision-related rights and duties in order to be able to adequately handle data as a company asset. |
| **Alhassan et al. (2016)** | Agreed with Otto's definition of data governance. Further explained that the definition of data governance indicates who holds the decision rights and accountability regarding an enterprise's data assets. Therefore, the decision domains should be identified in order to assign the right responsibilities and duties. |
| **Brous et al. (2016)** | Data governance is the exercise of authority, control, and shared decision-making over the management of data assets. Data governance provides organizations with the ability to ensure that data and information are managed appropriately, providing the right people with the right information at the right time. |
| **DAMA International (2017)** | Data governance is defined as: the exercises of authority, control, and shared decision-making (planning, monitoring, and enforcement) over the management of data assets. The goals of data governance are: 1) Enable an organization to manage its data as an asset. 2) Define, approve, communicate, and implement principles, policies, procedures, metrics, tools, and responsibilities for data management. 3) Monitor and guide policy compliance, data usage, and management activities. |
| **Abraham et al. (2019)** | Data governance refers to the exercise of authority and control over the management of data. Data governance specifies a cross-functional framework for managing data as a strategic enterprise asset. In doing so, data governance specifies decision rights and accountabilities for an organization's decision-making about its data. Furthermore, data governance formalizes data policies, standards, and procedures and monitors compliance. |
| **The Data Governance Institute** | Data governance is a system of decision rights and accountabilities for information-related processes, executed according to agreed-upon models which describe who can take what actions with what information, and when, under what circumstances, using what methods. |

Synthesizing the various definitions, it is found that the basic content of DG needs to include the following three points: (1) Data as an asset: data is recognized as a strategic enterprise asset that requires careful management. (2) Cross-Functional Framework: DG is cross-functional, involving multiple stakeholders across the organization. (3) Assigning decision-related rights and accountabilities: DG assigns decision-related rights and accountabilities of managing data asset.

Addressing these three features, we define DG in this study as follows: *the systematic and cross-functional exercise of managing data as a strategic enterprise asset. It involves recognizing data's value, assigning decision-related rights, establishing accountabilities across multiple stakeholders, and executing according to agreed-upon guidelines. DG entails structured authority and control over data, emphasizing formalized policies, procedures, principles, and monitoring compliance.*

**2.2 Distinction and Connection between DG and DM**

DG is often used interchangeably with DM. It is therefore necessary to clarify the differences and linkages between the two. DM consists of practices, architectural techniques, and tools for achieving consistent access and delivery of data across the spectrum of data subject areas and structure types in the enterprise (Gartner, 2022). DG and DM are highly interrelated, and scholars have discussed the difference between them. For example, some scholars believe that the main difference between the terms 'governance' and 'management' is that governance refers to the decisions that must be made and who makes these decisions in order to ensure effective management and use of resources, whereas management involves implementing decisions (Fu, Wojak, Neagu, Ridley, & Travis, 2011; Khatri & Brown, 2010). In this study, building upon the previously established definition of DG, DM is characterized as the exercise of "executing according to agreed-upon guidelines." Consequently, DM is regarded as an integral component of DG.

**2.3 Existing literature reviews on DG**

To comprehend the current state of the art, we have examined existing literature reviews on DG in a general or business context, aligning with our research focus. Following this screening process, we identified and retained 10 relevant literature reviews. Among these, 8 out of 10 were from the IS domain.

In 2016, Brous et al. confined their focus to the IS domain and argued that existing research has predominantly focused on DG structures, with limited attention given to the underlying principles.

They derived four principles for data governance, emphasizing organization, alignment, compliance, and common understanding.

Alhassan et. al. (2016) aimed to provide a comprehensive analysis of the activities involved in DG. They searched six major academic databases in the IS domain and identified 110 DG activities across five decision domains adopted from Khatri & Brown (2010), which are data quality, data principles, meta data, data access, and data life cycle. They defined DG as the combination of 'action,' 'area of governance,' and 'decision domain', with three action types: define, implement, and monitor. In 2018, they compared DG literature in scientific and practice-oriented publications, revealing that practice-oriented sources emphasized "implement" and "monitor" actions, while scientific publications focused more on defining activities.

Also hailing from the IS domain, Al-Ruithe et al. (2017) specialized in exploring DG within the context of cloud computing technology. They advocated for a comprehensive understanding of DG and its application in the realm of cloud computing. Their efforts focused on identifying potential common barriers and critical success factors for the implementation of cloud DG. In 2018, they introduced a taxonomy approach that encompassed both cloud and non-cloud DG, aiming to define distinct attributes and elucidate the concepts of DG in comparison to other governance domains. In 2019, they conducted a systematic literature review to gain insights into the state-of-the-art of DG in both non-cloud and cloud computing scenarios. This review enabled them to highlight challenges associated with implementing DG in cloud computing and identify key dimensions of DG.

Still within the IS domain, Merkus et al. (2019) sought to establish a unified set of definitions for DG and information governance within and across organizations, particularly in relation to the fundamental concepts of data and information. Drawing from existing literature on the Data-Information-Knowledge-Wisdom pyramid and discussions among peers, they classified and coded elements to create a new vocabulary. This laid the groundwork for developing a cohesive set of original definitions for data, information, meaning, DG, and information governance. Abraham et al. (2019) contended that despite the increasing importance of DG, a holistic view is lacking. Their study aimed to bridge this gap by developing a conceptual framework for DG. They identified the major building blocks of DG and broke them down along six dimensions: governance mechanisms, organizational scope, data scope, domain scope, antecedents, and consequences.

Two other works did not explicitly state their research domains, however, the frameworks they referenced primarily belong to the IS domain as well. Begg & Caira (2011) delved into the perceptions of small to medium enterprises (SMEs) regarding DG. They investigated the adaptability of current DG frameworks to SMEs and found that despite claims of adaptability and scalability, there was limited published evidence on the application of DG to SMEs. Additionally, they highlighted that effective utilization of DG frameworks requires individuals with authority and responsibility over enterprise data to possess knowledge and understanding of relevant terminology, which may not be met in many SMEs. Janssen et al. (2020) examined challenges and strategies associated with DG for Big Data Algorithmic Systems (BDAS). They put forth a DG framework to enhance the trustworthiness of BDAS. This framework advocates for stewardship of data, processes, and algorithms, controlled transparency for external scrutiny, trusted information sharing across organizations, risk-based governance, system-level controls, and data control through shared ownership and self-sovereign identities.

As observed, the majority of selected existing literature reviews are situated within the IS domain, primarily focusing on foundational aspects. Most of them are in the initial stages of comprehending DG, exploring underlying principles, key dimensions, definitions, conceptual frameworks, and DG activities. Only few of the studies have progressed beyond theoretical concepts, delving into the practical implementation of DG, e.g., examining common barriers and critical success factors for the implementation of cloud DG (Al-Ruithe et al., 2017), and challenges and strategies associated with DG for BDAS (Janssen et al., 2020).

Although these studies provide a solid theoretical foundation regarding principles, dimensions and concepts of DG, there is a noticeable gap in the literature reviews concerning the practical applications of DG within industry operations. This gap encompasses e.g., domains and scenarios where DG can be effectively deployed, the practical issues it can resolve, the specific difficulties hindering its implementation, and concrete examples illustrating its application. This gap arises from a lack of focus on DG within the realm of OSCM. Thus, our research endeavors to fill the gap by investigating the practical aspects of DG within OSCM, thereby enhancing our understanding of its real-world implications and potential benefits.

## 3. Methodology

### 3.1 Search Strategy

In conducting the literature review, a systematic methodology was employed to identify and select potential studies. The search was executed on the Web of Science database and Scopus database, limiting articles to those published in English and classified as journal articles. The selection of journals was guided by the Academic Journal Guide (ABS Ranking), a reputable source offering insight into journals relevant to business and management studies. For this study, we focused on 4*, 4, and 3 rated journals in the ABS Ranking in the fields of Operations and Technology Management and Operations Research and Management Science, amounting to a total of 39 selected journals. The ABS ranking's rigorous evaluation process, based on peer reviews, editorial assessments, and expert judgments, ensures the inclusion of high-quality and original research for our study. A detailed list of the selected journals is presented in Appendix 1.

Initially, we employed "data governance" as the search keyword in selected journals, yielding only a few relevant articles. Recognizing this inadequacy for a systematic literature review, we refined our approach. We leveraged existing literature reviews mentioned in Section 2.3 to reference the search strings employed by other researchers. The specific search terms identified are detailed in Table 2. We discovered that in addition to "data governance", scholars also utilized "governance data" (Al-Ruithe et al., 2018; Al-Ruithe et al., 2019) and "information governance" (Alhassan et al., 2016; Alhassan et al., 2018; Merkus et al., 2019; Abraham et al., 2019) as keywords for searching. In addition, as mentioned in Section 2.2, DM is often used interchangeably with DG. Thus, we decided to include "data management" in the search criteria. Ultimately, we settled on utilizing the keyword string: "data governance" OR "governance data" OR "information governance" OR "data management" for our search.

Table 2: Search strings employed by selected literature reviews

| Authors (year) | Title | Search terms |
|---|---|---|
| 1. Brous et al. (2016) | Coordinating decision-making in data management activities: a systematic review of data governance principles | "data governance" AND "principles" |
| 2. Alhassan et al. (2016) | Data governance activities: an analysis of the literature | "data governance" OR "information goverance" |
| 3. Alhassan et al. (2018) | Data governance activities: A comparison between scientific and practice-oriented literature | "data governance" OR "information goverance" |
| 4. Al-Ruithe et al. (2017) | Analysis and classification of barriers and critical success factors for implementing a cloud data governance strategy | - |
| 5. Al-Ruithe et al. (2018) | Data governance taxonomy: Cloud versus non-cloud | ((data governance) OR (data governance |

| | | | organization) OR (governance data) OR (data governance in cloud computing) OR (data governance for cloud computing) OR (cloud data governance)) |
|---|---|---|---|
| 6. | **Al-Ruithe et al. (2019)** | A systematic literature review of data governance and cloud data governance | ((data governance) OR (data governance organization) OR (governance data) OR (data governance in cloud computing) OR (data governance for cloud computing) OR (cloud data governance)) |
| 7. | **Merkus et al. (2019)** | Data Governance and Information Governance: Set of Definitions in Relation to Data and Information as Part of DIKW | "Data Information Knowledge Wisdom" OR "DIKW" OR "Data Governance" OR "Information Governance" |
| 8. | **Abraham et al. (2019)** | Data governance: A conceptual framework, structured review, and research agenda | "data governance" AND "information governance" |
| 9. | **Begg & Caira (2011)** | Data Governance in Practice: The SME Quandary Reflections on the Reality of Data Governance in the Small to Medium Enterprise (SME) Sector | - |
| 10. | **Janssen et al. (2020)** | Data governance: Organizing data for trustworthy Artificial Intelligence | - |

In our cumulative efforts, we amassed a total of 156 articles for consideration. Subsequently, we applied two-tiered selection criteria: 1) the research object is data, 2) the research scope is in the O&SC context. Following a meticulous screening process, we identified and retained a subset of 39 articles that met these criteria. The selection process of the study is presented in Figure 1.

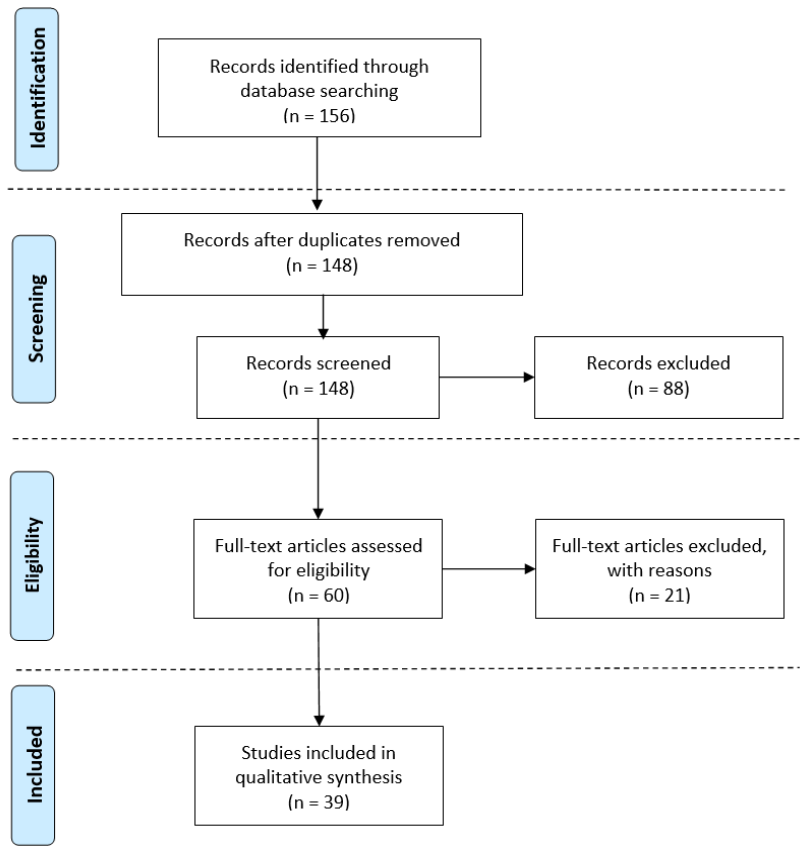

Figure 1. Flow chart describing the selection process of the studies included

(adapted from Moher et al., 2009)

**3.2 Statistical description of selected articles**

Understanding the landscape of selected articles is paramount to gaining insights into the prevailing trends, thematic concentrations, and scholarly contributions. This section provides a statistical overview, employing Figures 2, 3, and 4 to offer an examination of the selected literature.

Figure 2 traces the distribution of articles across different publication years, shedding light on the evolving patterns of research activity. Seen from the articles we collected, the interest in DG research within the OSCM domain has exhibited fluctuations over the three decades spanning from 1994 to 2023, with peaks observed in 2008 and 2014.

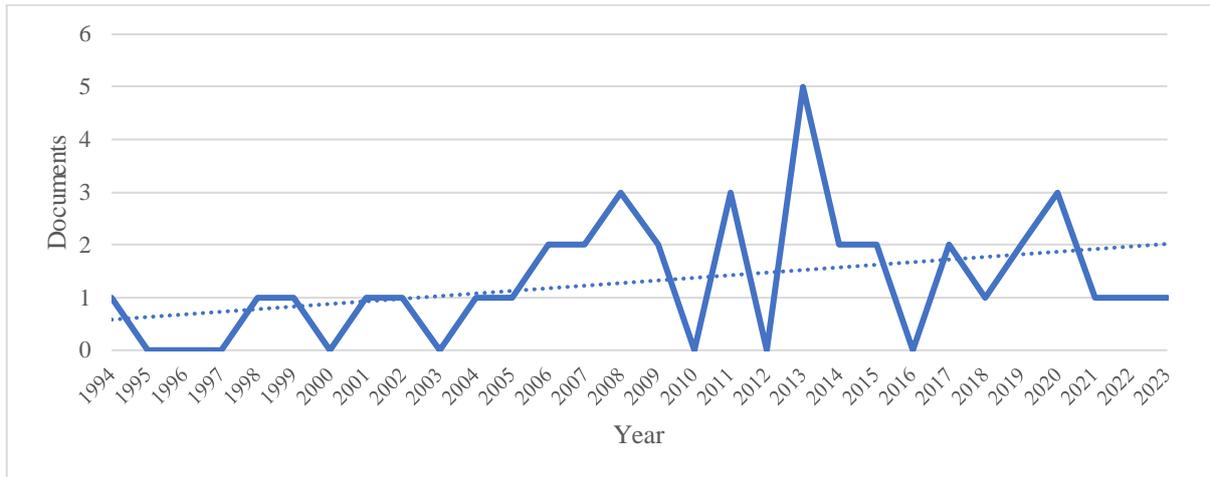

Figure 2. Documents by year

Figure 3 delves into the sources contributing to the body of literature, facilitating an assessment of the publication landscape. Among the surveyed journals, the majority of articles were sourced from *Computers in Industry* and *International Journal of Production Research*, accounting for 18 and 6 articles, respectively. This constitutes over 60 percent of the total collected articles.

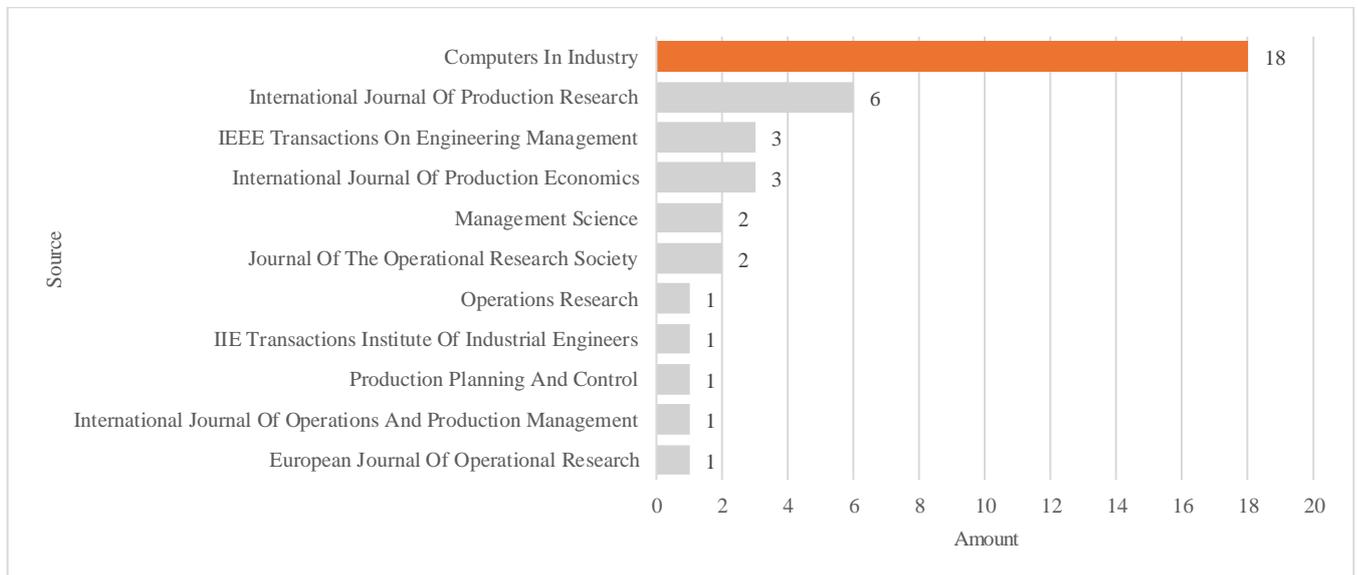

Figure 3. Documents by source

Lastly, Figure 4 categorizes articles by subject area, revealing the thematic diversity and concentration within the chosen literature. The primary emphasis of most articles lies within the areas of Engineering, Computer Science, and Business, Management and Accounting.

Together, these visual representations provide a comprehensive snapshot of the selected articles.

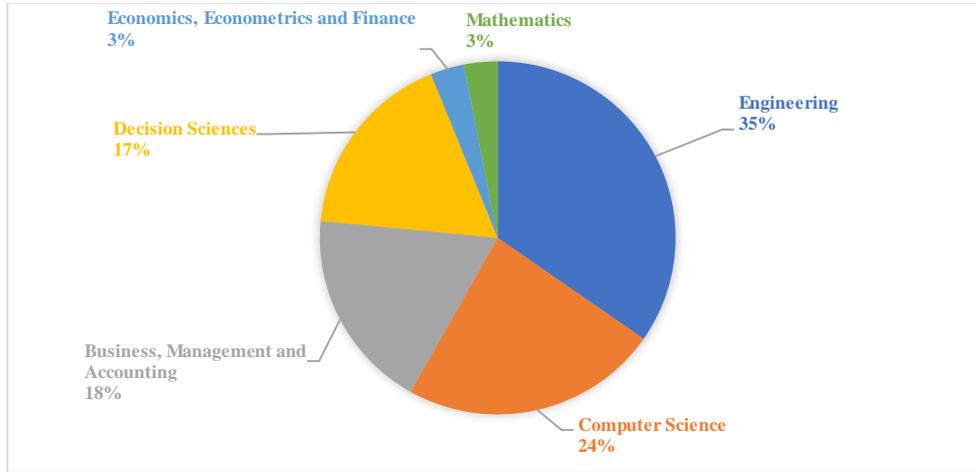

Figure 4. Documents by subject area

### 3.3. Article analysis framework

Our study aims to comprehensively explore the existing research on DG in the domain of OSCM. To achieve this, we analyzed the literature from both the OSCM and DG perspectives, and constructed an analytical framework grounded in these two perspectives.

*OSCM perspective*

From the OSCM perspective, our primary point of reference for conducting content analysis on the selected articles was the authoritative work titled "Operations Management" authored by Stevenson et al. (2007). This influential text enabled us to identify and categorize seven foundational facets within the domain of OSCM. These facets encompass *forecasting, system design, quality management, supply chain management (SCM), inventory management and scheduling, project management, waiting lines and simulation,* and *product development*.

Building upon the foundational facets of OSCM, we further classified these domains into three overarching categories: product development, manufacturing systems, and SCM. Product development, as a domain, encompasses elements such as *product design, product data management*, and *product configurations*. The manufacturing systems domain encompasses *forecasting, system design, quality management, inventory management and scheduling, project management,* and *waiting lines and simulation*. For the realm of SCM, we drew upon the influential work of Mentzer et al. (2001) as a point of reference. This domain was found to encompass various critical components, including *customer relationship management, customer service management, demand management, order fulfillment, manufacturing flow management, procurement,* among others. These elements collectively constitute the comprehensive landscape of OSCM within the context of our analysis.

*DG perspective*

DG is a broad concept and can be further unpacked into smaller elements. Specifically, in 2018, Saed et al. demonstrated the elements of an effective DG, namely, data architecture, classification and metadata, registration and audit information reports, data quality management, information lifecycle management, and security and information privacy. Alhassan et al. (2018) summarized five decision domains for DG (origin from IT governance), namely data principles, data quality, meta data, data access, and data lifecycle. Al-Ruithe et al. (2019) indicated that the domain scope of DG includes data quality, data security, data architecture, data lifecycle, meta data, data storage, and data infrastructure. Through synthesizing the studies, we identified the primary elements of DG as including *DM*, *data architecture, data lifecycle, data storage, data access, data quality, data security, metadata, data infrastructure,* and *data principles*.

When incorporating these elements into a content analysis framework, it is essential to consider their interrelationships and hierarchical structures. Among these elements, data lifecycle is a larger concept that consists various sub-components, such as data storage and data analysis. An influential article by Tao et al. (2018) delineates seven distinct phases within the manufacturing data lifecycle, comprising data sources, data collection, data storage, data processing, data visualization, data transmission, and data application. Notably, the data processing phase involves data cleaning, data reduction, data analysis, and data mining.

Therefore, we unfold the concept of "data lifecycle", and ultimately defined DG elements as follows: *DM, data architecture, data source and collection, data integration, data storage and access, data retrieval, data transmission and sharing, data mining, data analysis, data quality, data security, metadata, data infrastructure*, and *data principles*.

## 4. Analysis results

By applying the frameworks with two dimensions of OSCM and DG, this section classifies and analyzes the selected literature accordingly, and the statistical results are shown in Table 3. The results show that from the OSCM perspective, research related to data mainly focuses on product process (16 articles) and manufacturing system (15 articles), with relatively little research in SCM (6 articles). Additionally, two articles are dedicated to a general discourse on data, transcending specific domains delineated earlier. From the DG perspective, the research on data quality receives more attention (12 articles), followed by data transmission and sharing (5 articles), data storage and access (4 articles), data integration (4 articles), and data analysis (4 articles). The next sections will analyze these studies in detail.

Table 3. Statistical results of selected articles classification

|  | Product development | Manufacturing system | Supply chain management | Other work | Total |
|---|---|---|---|---|---|
| **Data management** | 1<br>Sackett & Bryan (1998) |  |  | 1<br>Ballou et al. (1994) | 2 |
| **Data architecture** | 1<br>Do & Chae (2011) |  |  |  | 1 |
| **Data source and collection** |  | 1<br>Guo et al. (2023) | 1<br>Spanaki et al. (2018) |  | 2 |
| **Data integration** |  | 4<br>Jiang (2007);<br>Hubel & Colquhoun (2001);<br>Rodrigues et al. (2008);<br>Espíndola et al. (2013) |  |  | 4 |
| **Data storage and access** | 1<br>Feng et al. (2009) | 3<br>Sen (2006);<br>Zong et al. (2017);<br>Villalobos et al. (2020) |  |  | 4 |
| **Data retrieval** | 2<br>Ou-Yang & Liu (1999);<br>Barton & Love (2005) |  |  |  | 2 |
| **Data transmission and sharing** | 3<br>Leong et al. (2002);<br>Qiu et al. (2007);<br>Low & Lee (2008) |  | 2<br>Shehab et al. (2013);<br>Epiphaniou et al. (2020) |  | 5 |
| **Data mining** |  |  |  | 1<br>Baesens et al. (2009) | 1 |
| **Data analysis** | 3<br>Do et al. (2014);<br>Do et al. (2015a);<br>Do et al. (2015b) | 1<br>Terrazas et al. (2019) |  |  | 4 |
| **Data application** |  | 1<br>Petrillo et al. (2019) |  |  | 1 |
| **Data quality** | 4<br>Parssian et al. (2004);<br>Li et al. (2013);<br>Tanaka & Kishinami (2006);<br>Son et al. (2010) | 5<br>Dey & Kumar (2013);<br>Chen et al. (2013);<br>Omri et al. (2021);<br>Hazen et al. (2017);<br>Schuster et al. (2022) | 3<br>Jalil et al. (2011);<br>Hazen et al. (2014);<br>Oliveira et al. (2019) |  | 12 |

| Data security | 1<br>Goel & Chen (2008) | | | | 1 |
|---|---|---|---|---|---|
| Total | 16 | 15 | 6 | 2 | 39 |

## 4.1 Data governance in the product development

*DM*

Product data management (PDM) is an important component of product development. PDM integrates and manages all the information that defines a product, from design to manufacturing, and to end-user support. However, in many engineering environments, PDM often proves to be suboptimal, manifesting issues related to data visibility, data accessibility, and data reliability (Sackett & Bryan, 1998).

*Data architecture*

In terms of data visibility, Do and Chae (2011) introduced a PDM architecture. This architecture, when augmented with integrated software configuration management, provides hardware engineers and software programmers with the capability to coexist harmoniously within a unified user environment. This environment operates on a cohesive database framework during the collaborative stages of product development.

*Data storage and access, Data retrieval, and Data transmission and sharing*

In terms of accessibility of data, many scholars have conducted research from the perspectives of data storage and access, data retrieval, data transmission and sharing. As far as data storage and access is concerned, Feng et al. (2009) introduced the concept of business components and proposed an integrated data management and storage framework for efficient heterogeneous design data management. As far as data retrieval is concerned, most of the PDM projects developed are based on the use of coding concepts to retrieve designs. However, the limited available codes constraint not only the flexibility of part shape representation, but also the versatility of retrieving existing drawings. Therefore, Ou-Yang & Liu (1999) proposed a comprehensive search method using topological relationship among form features, which is more convenient than the traditional code-based design retrieval methods. Based on a new version of an existing coding system (CAMAC), Barton & Love (2005) also developed a novel system that allows automatic coding of engineering drawings and subsequent retrieval using drawings of required components as the input.

As far as data transmission and sharing is concerned, Leong et al. (2002) proposed an object-oriented distributed database architecture capable of managing product data. This architecture facilitates the distribution of data to individuals who require it in a distributed manufacturing environment. Although secure and pervasive data sharing among all collaborative participants is critical to a successful PLM application, existing PDM solutions only provide rigid data access control mechanism like "all-or-nothing" scenario and thus prevent companies from sharing their data effectively. In response, Qiu et al. (2007) proposed role-based visualization to give enterprises and users data access rights of different details. Data from the product design, manufacturing and testing phases of the manufacturing industry are often either confused or unnecessarily repeated. Therefore, using templates to manage data at different stages of the project will not only realize the effective connection of data at each stage of the project, but also improve the responsiveness to changes in data as they occur (Low & Lee, 2008).

*Data analysis*

Considering that product data analysis can be used to evaluate and influence the product development process, Do (2014) mentioned that the application of Online Analytical Processing (OLAP) to a PDM database can effectively measure ongoing product development with the help of a set of process Key Performance Indicators (KPIs). Later, Do et al. (2015a) provided an integrated PDM database that can support both operational product data for product development and engineering change records for engineering change analysis. Do et al. (2015b) also proposed a product data analysis technology based on On-Line Analytical Mining (OLAM). In this approach, OLAM integrates On-line analytical processing (OLAP) with data mining, allowing it to support flexible and interactive OLAP operations to discover product data patterns in the invisible and uncertain product development process.

*Data quality*

In the context of ensuring data reliability, data quality emerges as a central and pivotal focus. Parssian et al. (2004) proposed that the first crucial step in managing data quality is to measure the quality of information products (derived data) based on the quality of source data and the related processes used to generate information outputs. Their research introduced a method to assess two data quality characteristics: accuracy and completeness. The method is generalizable and can be applied to determine how quality characteristics associated with different data sources impact the quality of derived data. In order to diagnose the data quality of geometric errors and avoid the cumbersome verification process in the receiving system, Tanaka & Kishinami (2006) also proposed a related quality criterion classification method and an authentication data model for product data. Their results showed that the diagnostic method is effective when applied to datasets including the error set provided by the Japan Automobile Manufacturers Association (JAMA). For the time delay and high cost of data correction caused by less reliable data, Son et al. (2010) developed a fully automated product data quality verification and management system that can effectively support the development process of high-tech products such as TVs, cameras, mobile phones and household appliances. In addition, Li et al. (2013) found that the PDM/PLM systems of enterprises usually lack superior mechanism to audit the information quality of product data and its impact on product design, manufacturing, service, and disposal. Therefore, they proposed a set of decisive auditing points and rules necessary for PDM/PLM systems, which may be a reference basis for companies to audit product data throughout the product lifecycle.

*Data security*

Finally, data security is of great significance to corporate strategic decision-making. It is argued that using a structured matrix-based risk analysis approach can effectively identify data security risks and link organizational assets at risk with security controls (Goel & Chen, 2008).

In short, current research on data in product development includes data architecture, data storage and access, data retrieval, data transmission and sharing, data analysis, data quality, and data security. A PDM architecture has been proposed for unified user environments for better data visibility. A data management and storage framework has been established for heterogeneous design data management. Data retrieval is not limited to using coding concepts, drawings using required components can also be used as input for subsequent retrieval. Data transmission and sharing is mainly based on role-based distribution and visualization to endow enterprises and users with data of different details, such as presenting model data with an appropriate level of detail reduction to users. There are relatively more studies on data quality and data analysis, mainly focusing on analyzing the impact of source data on the quality of derived data, proposing audit rules, diagnostic approach, or verification systems, and developing data analysis techniques to effectively evaluate and manage the ongoing product development process. Finally, research suggests that incorporating data security into strategic decision-making makes intuitive sense.

## 4.2 Data Governance in Manufacturing System

*Data source and collection*

In the era of smart manufacturing, distributed methods for data management and processing have garnered attention due to advantages such as a low cost of adaptation. However, the challenge remains in efficiently collecting data from various sources and discover proper range of data for smart decision-making. Guo et al. (2023) addressed this concern by proposing Dataspace as a viable and effective method to manage data from various sources. Dataspace facilitates the aggregation of distributed heterogeneous data from industrial enterprises.

*Data integration*

Data integration stands as another compelling area of focus which enhances data visibility by making data more accessible and coherent across various sources. However, discernible evidence within the capital plant industry underscores a persistent requirement for repetitive data entry at various stages of the bidding, design, and manufacturing processes. In addition, a noteworthy reliance on manual information processing persists, characterized by a lack of seamless integration among computer systems. Therefore, Hubel & Colquhoun (2001) introduced a new approach for capital plant manufacturing based on engineering data - Engineering Data Control (EDC) and explained the process flow and product structure of the EDC reference architecture to show how to integrate bids and orders processing flow. In addition, due to the heterogeneity of data from various sources, data integration is often one of the most challenging tasks in managing modern information systems. Jiang (2007) discussed a fundamental issue: when attribute values of real-world entities are recorded differently in different databases, how should one select the "best" value from the possible value sets? Accordingly, the author proposed a framework for integrating multiple data sources. Data transformation serves as a crucial step in the process of data integration. Most of organizations rely on eXtensible Markup Language (XML) standards to define data models. However, representing data in XML becomes problematic when different data sources need to be integrated. Therefore, Rodrigues et al. (2008) proposed a framework to allow organizations to automatically convert their XML data sources to semantic models defined in Web Ontology Language (OWL), thereby overcoming the above problems. Furthermore, for easier access and understanding of information from different systems, the CAx models integrated to Mixed Reality and Intelligent Maintenance (CARMMI) approach can support operators and technicians in maintenance tasks through mixed reality (Espíndola et al., 2013).

*Data storage and access*

When it comes to data storage and access, Sen (2006) explored the factors influencing the perception of Data Warehouse Process (DWP) maturity. This study firstly defined different levels of maturity for DWP and then conducted exploratory field research. The results indicate that several factors, including data quality, architectural consistency, change management, organizational readiness, and data warehouse size, impact DWP maturity. These findings provide useful management and technical insights for organizations looking to elevate their DWP to a more mature level. Another issue to consider is that ensuring data timeliness often comes with a high update cost, and immediately updating the database upon the arrival of new data is typically not the most efficient approach. Therefore, a crucial challenge lies in determining the optimal update strategy. By leveraging the Markov decision process model and solved through dynamic programming, the optimal update strategy can be obtained which minimizes the cumulative cost associated with both data outdatedness and updating (Zong et al., 2017). Furthermore, with a growing emphasis among manufacturers on monitoring and analyzing industrial systems, there is a concurrent rise in the expenses linked to storing the accumulated data. In addressing the imperative to manage and curtail these costs, Villalobos et al. (2020) advocated a three-level hierarchy for Industry 4.0 time-series data storage within cloud environments.

*Data analysis and Data application*

Data analysis has also been discussed. Considering that the vast amount of data generated by machines requires resources in terms of computing power and network bandwidth for analysis, Terrazas et al. (2019) proposed a novel big data approach and analytic framework for managing and analyzing machine-generated data in the cloud. The examined data serves diverse functions, prominently featuring its application in predictive modeling. A noteworthy illustration is found in the Automotive Predictor Cloud Maintenance (ATOMIC), where advanced big data technology is harnessed to formulate mathematical models delineating both nominal and faulty behaviors of automotive components. These models are instrumental in online estimation of end-of-life (EOL) and remaining useful life (RUL) indicators within the scrutinized automotive system, as outlined by Petrillo et al. (2019).

*Data quality*

Data quality has been drawn more attention in manufacturing system research. Dey & Kumar (2013) stated that prior research had explored how to estimate the quality level of database query outputs based on the quality level of input data. They summarized this research stream, allowing queries to have general selection conditions involving multiple attributes, as well as any combination of connected or separated sub-conditions that might include multiple attribute functions. Their results can be readily implemented in real-world decision-making environments. Chen et al. (2013) introduced a novel method for assessing and enhancing data quality in system health diagnosis modeling, using real-world industrial bearing test data as an illustrative example. Similarly, while data-driven prognostics and health management (PHM) can effectively serve as an asset performance management framework for data management and knowledge extraction, the issues with data quality often hinder the PHM process. In light of this, Omri et al. (2021) put forth a set of data quality criteria, particularly focusing on the relevance to fault detection tasks within PHM applications. After the methods for improving data quality had been discussed and gradually matured, some studies began to apply data quality improvement methods and evaluate the results. For example, Hazen et al. (2017) conducted a longitudinal single case study at an organization that maintains a large fleet of aircraft, collecting and analyzing qualitative interviews and observations, in order to explore outcomes that arise from a data quality improvement process implementation. Their findings indicate both positive and cautionary outcomes from implementing data quality improvement processes. These include heightened stakeholder commitment to data quality and business analytics, along with a potential overemphasis on program metrics at the expense of operational outcomes. Schuster et al. (2022) studied data quality in process mining. Their research pointed out that process mining provides methods and techniques for systematically analyzing event data. However, due to data quality issues and incomplete capture of process behavior, existing algorithms often produce low-quality models from real-life event data. Therefore, a new family of discovery algorithms has been developed that utilizes domain knowledge about the process in addition to event data. To organize this research, they presented a literature review of process discovery approaches exploiting domain knowledge.

In summary, the research on data in manufacturing systems is relatively extensive, mainly including data source and collection, data integration, data storage and access, data analysis, data application, and data quality. Among them, existing studies have paid more attention to data integration, data storage and access, and data quality. Research on data integration mainly focuses on how to effectively convert and integrate data to facilitate easy access and comprehension of information from various systems for data users. Research on data storage and access focuses more on elevating DWP maturity and reducing the cost of data storage and updating. Research on data quality touches upon several topics, including the estimation of database query output quality based on input data quality, exploration of outcomes resulting from the implementation of data quality improvement processes, methodologies for evaluating and enhancing data quality within the PHM domain, and the examination of data quality issues in process mining.

**4.3 Data Governance in Supply Chain Management**

*Data source and collection*

In the digital economy, the volume, variety, and availability of myriad forms of data from different sources has become an important resource for competitive advantage, innovation opportunities, as well as a source of new management challenges. Therefore, based on the theoretical and empirical foundations of traditional manufacturing supply chains, Spanalki et al. (2018) proposed the concept of data supply chain (DSC). In DSC, data is the main artifact flow. Their research outlines the characteristics of DSC, brings conceptual uniqueness to the context of DSC, and introduces innovative perspectives and pressing managerial dilemmas. The research combines conventional supply chain studies with data studies, offering a comprehensive exploration of this evolving field.

*Data transmission and sharing*

Other data research within the SCM domain is distributed across data transmission and sharing, and data quality. In terms of data transmission and sharing, to facilitate and strengthen product information sharing between original equipment manufacturers (OEMs) and their tier one suppliers, Shehab et al. (2013) introduced the development of four scenarios and designed a pull-mode industrial solution. The solution enables tier-one suppliers to retrieve data securely and efficiently from OEMs for manufacturing purposes, shortening product development cycles and improving the quality of manufactured components. Moreover, a present challenge lies in the electronic regulation of data sharing within and between diverse organizational entities in a supply chain. To address this, Epiphaniou et al. (2020) introduced Cydon, a decentralized data management platform. Cydon ensures authorized and rapid access to secure distributed data, mitigating single points of failure and preserving a continuous "always-on" chain of custody.

*Data quality*

Contemporary supply chain professionals contend with an abundance of data, and the efficacy of supply chain operations and planning hinges significantly on the quality of this data. Jalil et al. (2011) underscored the potential economic value embedded in installed base data for optimizing spare parts logistics. Furthermore, they delved into various data quality challenges linked with the utilization of installed base data, emphasizing the critical role of data quality dimensions in determining planning performance. Meanwhile, Hazen et al. (2014) scrutinized data quality concerns within SCM and provided valuable insights into methods for overseeing and regulating data quality. Finally, Oliveira et al. (2019) applied the organizational information processing theory to explore the intricate interplay among analytical capabilities, data quality, reporting quality, and the impact of real-time data capabilities on supply chain performance. Their research model posits that the benefits of real-time information technology hinge on the reporting quality and the effective management of analytical capabilities, collectively contributing to the enhancement of supply chain performance.

In a word, DG in the field of SCM mainly focuses on data source and collection, data transmission and sharing, and data quality. To manage heterogeneous-source data, a novel concept known as DSC has emerged, signifying a fusion of research in the domains of data and SCM. In term of data transmission and sharing, the main focus is to optimize secure access to information within an enterprise or among upstream and downstream manufacturers in the supply chain, which aims to enhance product quality and streamline the product cycle. Given the significant influence of data quality on supply chain efficiency and strategic planning, current research predominantly focuses on methods for detecting and controlling data quality.

**4.4 Data governance in others**

There are also two studies, i.e., Ballou et al. (1994) and Baesens et al. (2009), focusing on data in the realm of OSCM. Instead of being classified into specific domains of product development, manufacturing systems, or SCM, these two studies encompass research that holds broader applicability across various contexts related to data. They are distributed across the topics of data management and data mining.

On the one hand, Ballou et al. (1994) explored data management project. They argued that facing with increasingly diverse data resource, enterprises should firstly determine the priorities of projects that affect or involve various aspects of data resources, and then manage relevant data effectively. They also proposed a multi-stage process designed to accommodate the ambiguity inherent in assessing the benefits that will be derived from a data management project. On the other hand, Baesens et al. (2009) focused on the topic of data mining. They argued that data mining plays an increasingly important role in decision-making. However, there were still many challenges to be addressed, from data quality issues to how to incorporate domain experts' knowledge or how to monitor model performance. In their study, they provided an overview of upcoming trends and challenges in data mining and its role in operations research.

## 5. Future research

*5.1 Product Development*

In the context of product development, the identified literature lacks research on data integration and data application. Greater emphasis should be placed on these areas, particularly on data integration. This is crucial given the contemporary diversity in the representation of product data, encompassing various formats such as text, images, videos, etc.

Barton and Love (2005) underscored the utilization of drawings data for retrieval purposes. However, research on transforming and integrating data with different formats remains inadequate. With the advent of the big data era, the landscape of data has evolved significantly, characterized by increased diversity, voluminous quantities, and a surge in unstructured and non-textual data in the product design domain. Consequently, it becomes imperative to shift focus towards comprehensive research on data integration methodologies.

Future research can focus on data scalability in the data integration, investigating what strategies and architectures are effective in managing large-scale data in complex product development environments. Furthermore, scholars can explore real-time data integration by examining approaches and technologies that facilitate its implementation in product development, which enable immediate access to updated information throughout the entire product lifecycle. It is also worth studying the development of interconnected ecosystems. This involves addressing how data integration strategies can be adapted to accommodate the increasing interconnectedness of product development ecosystems, involving collaborations with external partners, suppliers, and customers. Such attentions are warranted to enhance the efficacy of DM and DG in the product development domain.

*5.2 Manufacturing Systems*

In the context of manufacturing systems, the exploration of data retrieval, data transmission, data transmission and sharing, and data security remains conspicuously absent in the collated literature. Among these, particular emphasis should be directed toward research endeavors addressing data transmission and sharing, and data security, given their critical significance in contemporary manufacturing landscapes.

Research on data transmission and sharing in manufacturing systems is vital for enhancing operational efficiency, collaboration, and adaptability. In the context of Industry 4.0, data transmission and sharing are foundational for implementing smart manufacturing technologies, including the Internet of Things, artificial intelligence, etc. In interconnected manufacturing processes, data transmission and sharing enables seamless communication, real-time decision-making, and resource optimization. Within the industry, the concern regarding data transmission and sharing touches upon both technical considerations and the willingness of sharing data. Therefore, future research should explore the application of edge computing to enable real-time data transmission and sharing, addressing scalability challenges, and optimizing data transmission and sharing through for example machine learning. Additionally, research should investigate security and privacy measures and interoperability standards, e.g., the integration of blockchain technology for enhanced transparency and traceability, as well as focusing on the development of governance frameworks and ethical considerations for cross-organizational collaboration and ensuring responsible data use. Finally, understanding regulatory compliance challenges and navigating evolving data protection regulations will be essential for manufacturers in developing effective data transmission and sharing strategies.

*5.3 Supply Chain Management*

In the context of SCM, research on data is limited. Existing studies have predominantly centered on data transmission and sharing, and data quality. However, considering their broad relevance and impact, data integration, data analysis, and data security are considered crucial as well.

Some researchers have noticed that effective research on data integration is crucial in SCM due to the complex nature of supply chains (Vieira et. al., 2020), which involve diverse data sources and interconnected processes. Future research can focus on flexibility, delving into strategies that enhance the adaptability of data integration systems to evolving supply chain dynamics. Additional emphasis should also be placed on cross-organizational integration, discussing how can trust and collaboration be fostered among diverse stakeholders for successful data transmission and sharing, and data integration.

Data analysis is indispensable in supply chain management, as it enables informed decision-making to optimize operations, streamline inventory management, and enhance resource allocation. With the development of advanced predictive analytics, future research could focus on exploring how can advanced predictive analytics techniques, such as Machine Learning (ML) and Artificial Intelligence (AI), be applied to enhance performance in the SCM domain. Human-centric data analytics is another intriguing topic to investigate. Future research can explore how can human-centric data analytics help understand user behaviors, preferences, and decision-making processes within the supply chain.

Robust data security measures are essential for preventing disruptions, fostering trust among stakeholders, and maintaining the confidentiality and integrity of shared information, ultimately contributing to a secure, resilient, and trustworthy SCM ecosystem. Future research can further explore the application of blockchain technology in supply chain security, discussing what role can blockchain technology play in enhancing the security of supply chain data, and how can its adoption be optimized to ensure secure and transparent transactions. Cross-border data security is also worth investigating, in order to understand the key challenges and solutions for ensuring data security in globalized supply chains. This involves a comprehensive examination of diverse regulatory environments and the complexities associated with cross-border data flows.

Finally, the DSC concept proposed by Spanalki et al. (2018) builds a solid foundation for studying data through the integration of supply chain knowledge. This serves as an exemplary illustration of disciplinary integration. Subsequent research endeavors should leverage this conceptual framework to explore data within the context of supply chain domain knowledge.

*5.4 General Data Research Suggestions*

In the realm of OSCM, three additional noteworthy points merit attention when studying data. First, there is the significance of studying data based on types. In our analysis of the literature, one specific study by Terrazas et al. (2019) delved into the management and analysis of machine-generated data. Recognizing that diverse data types necessitate distinct management methods is crucial; for instance, human-centric data markedly differ from machine-generated data in terms of velocity, volume, value, variety, and veracity. Hence, it is imperative to examine data based on various types.

Second, exploring how to manage data in different domains is vital. In our analysis, we found two studies focusing on data quality in the PHM domain (i.e., Chen et al., 2013; Omri et al., 2021). Similar with these two studies, more research that delves into specific domains are encouraged, as they can establish clear industry links and facilitate practical implementation in the real world.

Finally, in the midst of the rapid ascent and evolution of AI, the exploration of leveraging technologies such as ML, Deep Learning, and large language models to enhance business operations and SCM has garnered significant attention. Nevertheless, research concerning data in AI implementation is still in its developmental stages. Consequently, exploring methods to effectively prepare data for utilization in these technologies emerges as a highly promising research focus.

**6. Conclusion and limitation**

In this study, we defined DG as *the systematic and cross-functional exercise of managing data as a strategic enterprise asset. It involves recognizing data's value, assigning decision-related rights, establishing accountabilities across multiple stakeholders, and executing according to agreed-upon guidelines. DG entails structured authority and control over data, emphasizing formalized policies, procedures, principles, and monitoring compliance*. Subsequently, we delineated the constituent elements of DG. Building upon this foundation, we developed an analytical framework to scrutinize the collected literature from the perspectives of both OSCM and DG.

Through the systematic literature review, we found that in the realm of production development, research is concentrated on data quality and analysis. This includes proposing audit rules, diagnostic approaches, and verification systems for data quality, as well as developing data analysis techniques to manage ongoing product development effectively. Incorporating data security into strategic decision-making is also highlighted. In manufacturing systems, extensive research covers data management projects, integration, storage, quality, analysis, and application. Notably, attention is focused on data integration to facilitate access to information from different sources, cost-effective data storage, and methods for enhancing data quality to improve usability. In SCM, DG research emphasizes data transmission and sharing and data quality. Key objectives include ensuring secure data access within enterprises and the supply chain, enhancing product quality, and shortening product cycles. Most of the research centers on methods for detecting and controlling data quality. Additionally, a novel concept, DSC, has emerged, amalgamating data and supply chain research.

Furthermore, some studies that discuss data from a broader perspective also encompass topics like data integration, data storage and access, data quality, and data mining, with the primary focus of research concentrated on data quality.

The contributions of this study are twofold. For academia, the contribution lies in filling a significant gap. By compiling and summarizing DG-related research in OSCM, the study provides scholars and researchers with a consolidated and comprehensive view on the relevant literature. It enhances the understanding of DG within the unique context of OSCM, offering insights that can inform further academic exploration and inquiry. For industry, the contribution lies in the practical relevance of the research findings. The insights derived from the research can be directly applicable to real-world scenarios within O&SC contexts. This facilitates the creation of practical guidelines and approaches tailored for organizations helping them enhance their DG practices and derive maximum value from their data assets.

While this study provides valuable insights into redefining DG, outlining its constituent elements, and developing an analytical framework, the limitation of this study is that it may not explore methodologies that could enrich the understanding of DG in OSCM. Meanwhile, the retrospective analysis approach adopted in the study may limit its ability to capture emerging trends or dynamic changes in the field. Additionally, this study may not thoroughly explore the practical challenges and implementation issues faced by organizations in adopting effective DG practices. Therefore, further research could explore these aspects and provide a more comprehensive understanding of DG's role in contemporary business environments.

**Acknowledgement:** This study is supported by MADE FAST (MADE -- Flexible, Agile, and Sustainable production enabled by Talented employees) Project (470100), Jiangxi Double Thousand Plan, and "Strengthening the digitalization of businesses in Eastern Europe – a micro and macro-level approach" funded by the European Union – NextGenerationEU project and the Romanian Government.

**Disclosure statement:** The authors report there are no competing interests to declare.

**Biographical note:**

**Xuejiao Li**

Ph.D. candidate at Aalborg University, Denmark, with a focus on operations and supply chain management, specifically data governance in the manufacturing industry. Her research aims to develop data governance frameworks, guidelines, and methods to help companies achieve higher performance by capturing the value in their data assets. Her work is highly related to digital, data and analytics, and operations.

**Yang Cheng**

Associate professor in Center for Industrial Production, Aalborg University, Denmark. Master of Management from Beihang University, China and a PhD from Aalborg University. Cheng has extensive research and consulting experience in business process reengineering, manufacturing strategy, global manufacturing network management and knoweldge transfer. Specialties: Business process reengineering, manufacturing/operations strategy, global manufacturing/operation network management, and knowledge transfer.

**Charles Møller**

Professor at Department of Mechanical and Production Engineering, Aarhus University, Denmark. Research and teaching: Supply Chain Management, Enterprise Systems Management, Business Process Management. Specialties: business process innovation.

**Xiaoning Xia**

Visiting Ph.D. student at Aalborg University, Denmark. Ph.D. candidate at Chongqing University, China. Specialties: circular economy, energy economy, and life cycle assessment.

Zong, W., Wu, F., & Jiang, Z. (2017). A Markov-based update policy for constantly changing database systems. *IEEE Transactions on Engineering Management*, *64*(3), 287-300.

**Appendix**

Appendix 1. Selected journals

| Field | Journal Title | ABS Ranking |
|---|---|---|
| Operations and technology management (12) | Journal of Operations Management | 4* |
| | International Journal of Operations and Production Management | 4 |
| | Production and Operations Management | 4 |
| | Computers in Industry | 3 |
| | IEEE Transactions on Engineering Management | 3 |
| | International Journal of Production Economics | 3 |
| | International Journal of Production Research | 3 |
| | Journal of Scheduling | 3 |
| | Journal of Supply Chain Management | 3 |
| | Manufacturing and Service Operations Management | 3 |
| | Production Planning and Control | 3 |
| | Supply Chain Management: An International Journal | 3 |
| Operations research and management science (27) | Management Science | 4 |
| | Operations Research | 4 |
| | European Journal of Operational Research | 4 |
| | IEEE Transactions on Evolutionary Computation | 4 |

| | |
|---|---|
| Mathematical Programming | 4 |
| IEEE Transactions on Systems, Man, and Cybernetics: Systems (formerly "IEEE Transactions on Systems, Man and Cybernetics - Part A: Systems and Humans") | 3 |
| ACM Transactions on Modeling and Computer Simulation | 3 |
| Annals of Operations Research | 3 |
| Computational Optimization and Applications | 3 |
| Computers and Operations Research | 3 |
| Decision Sciences | 3 |
| Evolutionary Computation | 3 |
| Fuzzy Optimization and Decision Making | 3 |
| IEEE Transactions on Cybernetics (formerly "IEEE Transactions on Systems Man and Cybernetics Part C (Applications and Reviews)") | 3 |
| IIE Transactions | 3 |
| INFORMS Journal on Computing | 3 |
| International Journal of Forecasting | 3 |
| Journal of Heuristics | 3 |
| Journal of Optimization Theory and Applications | 3 |
| Journal of the Operational Research Society | 3 |
| Mathematics of Operations Research | 3 |
| Naval Research Logistics | 3 |
| Omega: The International Journal of Management Science | 3 |
| OR Spectrum | 3 |
| Reliability Engineering and System Safety | 3 |
| SIAM Journal on Optimization | 3 |
| Transportation Science | 3 |